# How Does van der Waals Confinement Enhance Phonon Transport?

Xiaoxiang Yu[1,2,#], Dengke Ma[1,2,#], Chengcheng Deng[3,*], Xiao Wan[1,2], Meng An[1,4], Han Meng[1,2], Xiaobo Li[3], Xiaoming Huang[3], Nuo Yang[1,2,*]

[1] State Key Laboratory of Coal Combustion, Huazhong University of Science and Technology, Wuhan 430074, P. R. China

[2] Nano Interface Center for Energy (NICE), School of Energy and Power Engineering, Huazhong University of Science and Technology, Wuhan 430074, P. R. China

[3] School of Energy and Power Engineering, Huazhong University of Science and Technology, Wuhan 430074, P. R. China

[4] College of Mechanical and Electrical Engineering, Shaanxi University of Science and Technology, Xi'an 710021, China

# X. Y. and D. M. contributed equally to this work.

* To whom correspondence should be addressed. E-mail: nuo@hust.edu.cn (N.Y.), dengcc@hust.edu.cn (C.D.)

# Abstract

The van der Waals (vdW) interactions exist in reality universally and play an important role in physics. Here, we show the study on the mechanism of vdW interactions on phonon transport in atomic scale, which would boost developments in heat management and energy conversion. Commonly, the vdW interactions are regarded as a hindrance in phonon transport. Here, we propose that the vdW confinement will enhance phonon transport. Through molecular dynamics simulations, it shows that the vdW confinement makes more than two-fold enhancement on thermal conductivity of both polyethylene single chain and graphene nanoribbon. The quantitative analyses of morphology, local vdW potential energy and dynamical properties are carried out to reveal the underlying physical mechanism. It is found that the confined vdW potential barriers reduce the atomic thermal displacement magnitudes, thus lead to less phonon scattering and facilitate thermal transport. Our study offers a new strategy to modulate the heat transport.

The van der Waals (vdW) interactions, which arise from the electromagnetic forces between quantum fluctuation-induced charges [1], exist in reality universally. The vdW interactions play an significant role in fields as diverse as condensed matter physics [2-6], reaction's dynamics in chemistry [7] and assembly of complex supramolecule in biology [8]. In the area of heat transfer, the vdW interactions can remarkably impact phonon transport in nanostructured materials due to the large surface-to-volume ratio [9, 10]. Phonon transport in nanoscale is of great importance to many practical applications, including energy conversion and package of electronics [11-15]. Generally, nanomaterials are utilized in specific forms like in-bundles or with-substrates, where there are inevitable vdW interactions. Hence, there is an intense demand to understand how vdW interactions affect phonon transport in nanostructured materials.

Tremendous efforts have been devoted to the investigation of this topic in past decades [16]. Commonly, theoretical and experimental studies concluded that vdW interactions block phonon transport and lead to a reduction of thermal conductivity (TC) in nanostructures [17-23]. Recently, Yang *et al.* and Qian *et al.* experimentally demonstrated an enhancement of TC by vdW interactions for graphene-like h-BN nanoribbon [24] and quasi-1D van der Waals crystal $Ta_2Pd_3Se_8$ nanowires [25], respectively. In understanding of its effect on phonon transport, there are only studies in the intensity of vdW interactions by simulating Frenkel-Kontorova lattices [26] and Fermi-Pasta-Ulam chains [27].

In this letter, it studies theoretically how vdW interactions modulate, especially enhance, the phonon transport in nanoscale. Firstly, a strategy on artificially designing the nanostructures is proposed. Secondly, based on equilibrium molecular dynamics (EMD) simulations on typical nanostructures designed by the strategy, it shows that the vdW interactions may enhance phonon transport. Lastly, to reveal the underlying physical mechanism, it is carried out that the quantitative analyses of morphology, local vdW potential energy and dynamical properties.

We proposed a new vdW confinement effect that the vdW interactions could form potential barriers on atoms, thus may confine the vibrational amplitude, which will decrease anharmonic phonon scattering and facilitate thermal transport. The vdW potential barriers can be made by artificially designing the nanostructures. So, a proper designing strategy is the key issue. In the following, an alternative artificial structure, named as the crosswise paved laminate (CPL), is introduced. And, the thermal conductivities (TCs) of CPL structures are obtained by EMD simulations, which are compared with TCs of other corresponding structures, like single-chain polyethylene (SC-PE) [28] and single-layer graphene nanoribbon (SL-GNR) etc.

The CPL structures are constructed based on PE chains and GNRs, respectively (see details in section 1 of supplementary information (SI)). As depicted in Fig. 1(a-b), the aligned chains or ribbons are paved layer by layer, and the aligned directions are crosswise for two adjacent layers. The intersection angle of PE chains is 90 degree, while that of GNR is set to 120 degree due to ordered stacking of its hexagonal structure. It is worth noting that the CPL structures are probably fabricated (see details in section 2 of SI) according to a recent experimental research on the weaving of organic threads [29]. Also, there are already analogous structures reported, such as orthogonal self-assembly multilayer polymer meshes [30], ordered polymer structure [31].

The TCs of CPL structures was calculated by Green-Kubo formula [32] using LAMMPS package [33] (simulation details shown in section 3 of SI). Although the validation of Green-Kubo formula for low-dimensional systems remains inconclusive [34-36], there are also many studies on thermal transport in both nanostructure and bulk [37, 38]. We did not discuss this issue here. The inter-atomic interactions in PE and GNR are described by the adaptive intermolecular reactive bond order (AIREBO) potential [39-42], which provides a realistic description of the dynamics and anharmonicity [43, 44]. To accurately simulate an infinite system with a finite MD simulation cell, periodic boundary conditions are applied along all three directions, which inhibit phonons from being attenuated by boundaries. Length dependence of TC was investigated to overcome the size effect (see details in section 4 of SI). The TCs of CPL structures along two in-plane directions (x and y) are much higher than that along cross-plane direction (z) (< 1 $Wm^{-1}K^{-1}$) contributed by nonbonding interactions. So we only focus on the TCs of CPL structures along two in-plane directions throughout this article.

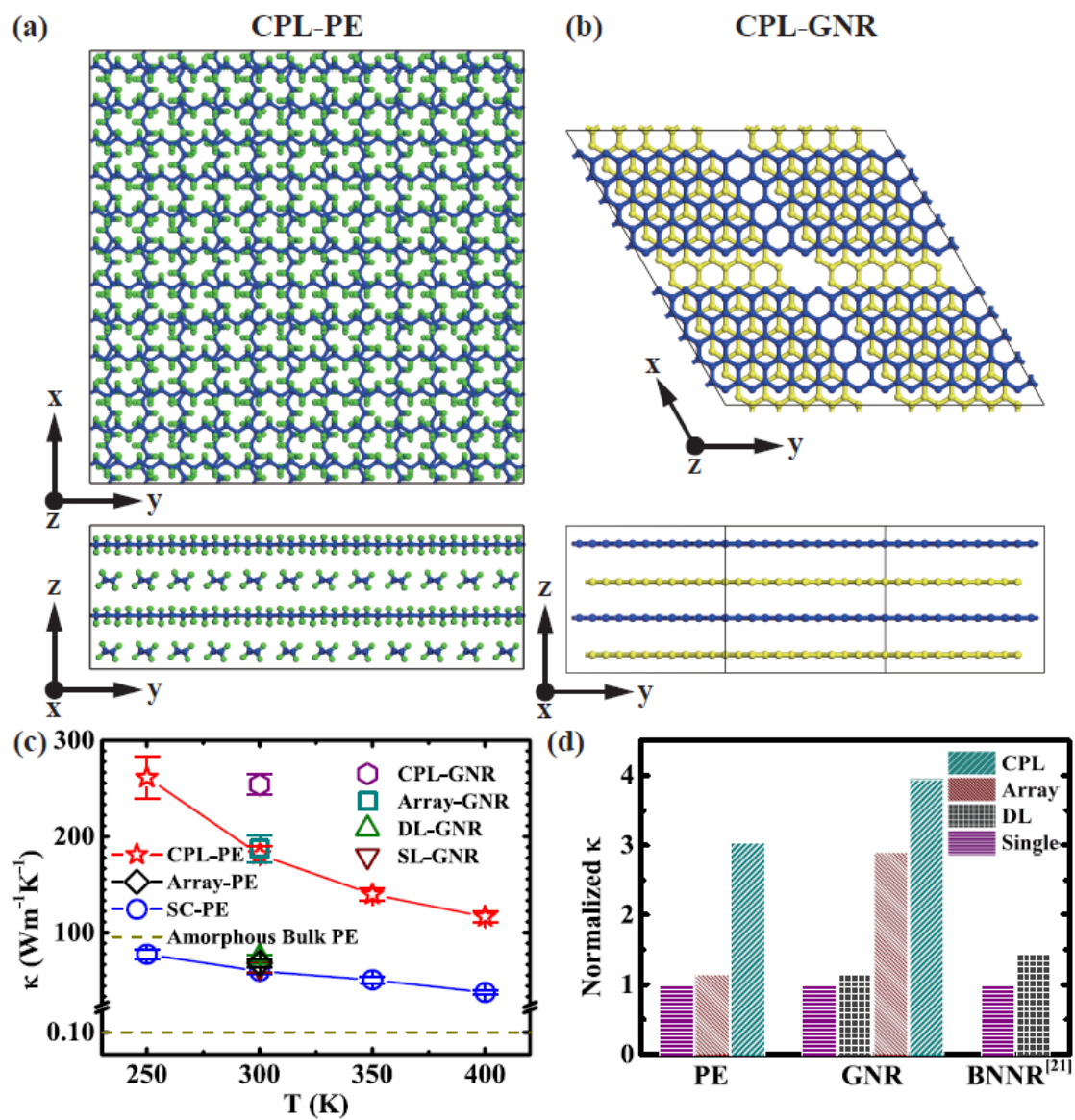

**FIG. 1.** (a-b) Schematic views of simulation cell of crosswise paved laminate (CPL) for polyethylene (PE) chain and graphene nanoribbon (GNR). (c) The TCs of PE chains and GNRs systems as functions of temperature. (d) The ratios of thermal conductivity for structures with (CPL, array and double layer (DL)) and without vdW interactions (single chain/ribbon).

As shown in Fig. 1 (c), the TC of SC-PE and SL-GNR are 54 Wm$^{-1}$K$^{-1}$ and 64 Wm$^{-1}$K$^{-1}$ at room temperature, respectively. The $\kappa_{SC-PE}$ is comparable to previous simulation studies [45, 46]. The value of $\kappa_{SL-GNR}$ is slightly smaller than the previous simulation result using same methods [20, 44, 47], which is ascribed to a smaller width and size confinement effect. With the effect of vdW interactions, the TCs of both PE and GNR systems gradually increase for arrays and CPL structures. Especially, the TCs of CPL structures for PE and GNR are several times larger than that of single one, due to vdW confinement which decreases phonon scattering as will be discussed later in detail. At room temperature, $\kappa_{CPL-PE}$ and $\kappa_{CPL-GNR}$ reach as high as 181 Wm$^{-1}$K$^{-1}$ and 254 Wm$^{-1}$K$^{-1}$ respectively. Both SC-PE and CPL-PE show negative temperature dependence of TC. Moreover, the values of $\kappa_{CPL-PE}$, such as 181 Wm$^{-1}$K$^{-1}$ at 300 K, are conservatively calculated using the whole bulk cross section area. When we consider the contribution of each layer to the thermal transport, the layers, where PE chains are perpendicular to direction of heat transfer, have an ultra-low TC. So, most of the contributions come from the layers where PE chains are parallel to the direction of heat transfer, whose TCs are estimated to be twice of $\kappa_{CPL-PE}$, such as ~ 362 Wm$^{-1}$K$^{-1}$ at 300 K. And the possible high TC of the CPL structure for GNR can be predicted in the similar way. Furthermore, the CPL structures have high TCs along two in-plane directions, adding an extra dimension for efficient heat dissipation. To some extent, high TC along two directions means one effective way of enhancement of thermal properties.

Moreover, due to vdW potential barriers, Fig. 1 (d) presents the enhancement ratios of TC for PE chains, GNRs and boron nitride nanoribbons (BNNRs), respectively. Compared with SC/SL structure, array-PE has slightly improvement of TC, while array-GNR shows almost two times increase in TC. Especially, CPL-PE and CPL-GNR respectively exhibit two times and nearly three times enhancement on TC. Besides, the TC of DL-GNR is higher than that of SL-GNR, which is consistent with experimental result of SL and DL graphene-like hexagonal boron nitride nanoribbon (BNNR) [24].

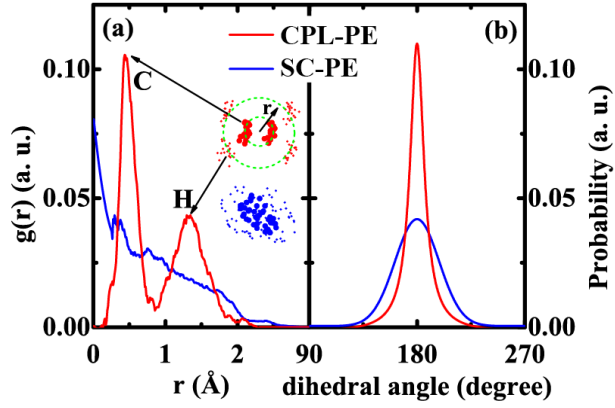

**FIG. 2.** (a) Radial distribution functions (RDFs) g(r) of carbon and hydrogen atoms and (b) probability distributions of dihedral angles of C-C-C-C backbone in PE chain.

To gain insights into the phonon transport enhancement, the mechanism of vdW confinement will be shown in the following. Herein, we only show the results about PE in the main body to avoid repeated description of similar results about GNR (shown in SI). The morphology of PE chains is quantified to exhibit atomic vibrations through radial distribution functions (RDFs) (calculation details given in section 5 of SI) and probability distribution of the dihedral angle of the C-C-C-C backbone. Fig. 2 (a) shows the significant differences of lattice orders between SC-PE and CPL-PE. For SC-PE, there is no obvious peak in RDFs. Correspondingly, the distribution along chain axis (blue dots) indicates that atomic vibrations in SC-PE are full of bending motions. In contrast, there are two sharp peaks in RDFs of one PE chain in CPL-PE. According to the distribution along chain axis (red dots), the two peaks in RDFs of CPL-PE correspond to the equilibrium positions where C atoms and H atoms are located (green circles). It denotes that CPL-PE has more crystal-like atomic vibrations of hydrogen atoms and subdued bending motion of backbone than SC-PE, which is conducive to a higher TC. Besides, from the probability distributions of the dihedral angles of the backbone as shown in Fig. 2 (b), there is a much steeper peak for one chain in CPL-PE than the SC-PE. That means, there is much weaker torsional motion of one PE chain in CPL-PE than in SC-PE. In brief, vdW confinement enables CPL-PE more crystal-like, so as to induce less phonon scattering, leading to the enhancement of thermal conductivity.

To elaborate the vdW confinement in CPL, we analyzed the potential energy barrier of both CPL-PE and array-PE (calculation details shown in section 6 of SI). The local potential energy barrier is mapped to evaluate the confinement on atomic vibrations by vdW interactions. For translational motion, there is an obvious confined vdW potential barrier in CPL-PE (Fig. 3 (a)), while the array-PE shows small potential change along y direction (Fig. 3 (b)). Comparing to CPL-PE, atoms in array-PE can move easily along y direction, thus there is less vdW confinement. This discrepancy stems from the unidirectional alignment of array-PE and crosswise layout of CPL-PE, which indicates CPL-PE could confine the atomic motions more effectively than array-PE does. CPL-GNR also shows confined potential energy barriers (given in section 7 of SI). Furthermore, Fig. 3 (c) demonstrated the comparison of potential energy changes of PE chain in CPL-PE and array-PE as a function of rotation angle. As can be seen, CPL-PE shows bilateral steeper potential barrier than array-PE. So CPL structure can confine the segmental rotation better, thus diminish anharmonic phonon scattering, which facilitates thermal transport.

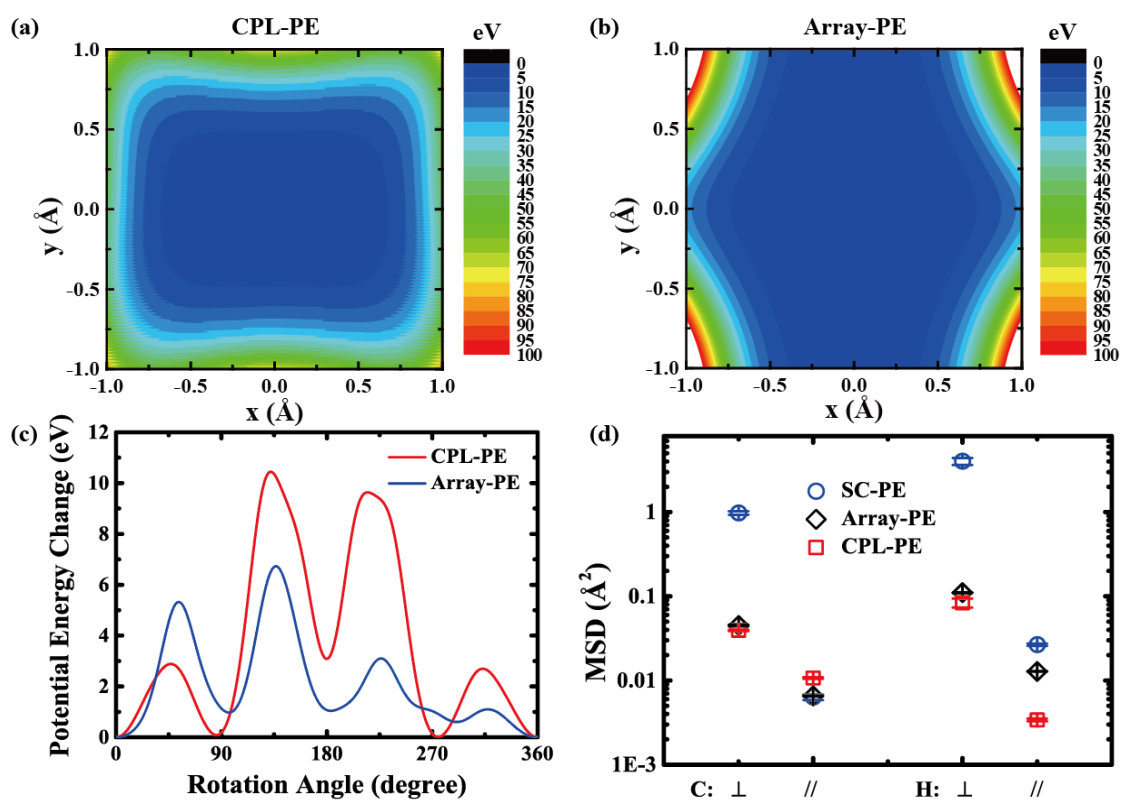

**FIG. 3.** The potential energy barriers of PE chain in (a) CPL-PE and (b) PE array. (c) The potential energy changes as a function of rotation angle. There is a confined vdW potential barrier in CPL-PE in comparison with PE array. (d) The mean square displacement (MSD) along radial ($\perp$) and axial ($//$) directions of carbon and hydrogen atoms.

To shed further light on the anharmonic properties, we calculated mean square displacement (MSD) through MD simulations including full anharmonicity by directly using time-averaging positions. Since the atoms are only allowed to displace from equilibrium through the natural energy of the thermal fluctuations, the closer the atom deviates from the equilibrium position, the less phonon-phonon scattering occurs which means the system is less anharmonic [48, 49]. As shown in Fig. 3 (d), the MSD of PE chains in CPL-PE, especially that of hydrogen atoms, is much smaller than of that of SC-PE. Recent theoretical work emphasized that the large thermal displacements of hydrogen strongly affect the motions of carbon atoms, thereby altering the scattering phase space and reducing the TC [49]. In our work, the thermal displacements of hydrogen atoms are reduced by vdW confinement and thus cause the TC enhanced.

Last but not least, we would like to provide a predictive framework, namely, two implications of applying our designing strategy to achieve vdW confinement. On the one hand, the preferable materials to achieve positive effect of vdW confinement on thermal transport would better have two characteristics: i) extremely large surface-to-volume ratio, like one-dimensional (1D) and quasi-1D nanostructures; ii) stiff along axial direction while soft along radial direction, namely, potentially good thermal transport along axial direction but strong phonon scattering by anharmonic radial atomic vibration. The first characteristic leads to considerable effect of vdW interactions. The second characteristic ensures the possibility to improve TC via vdW interactions. On the other hand, the preferable structures that are conducive to forming strong confined vdW potential barrier, are radially parallel, axially crosswise and periodically ordered layout e.g. CPL. The parallel structures enable the vdW barriers ordered thus confine the anharmonic atomic vibration by each other. Meanwhile, the crosswise structure is able to diminish the torsional motion better. Besides, periodically ordered structure could further strengthen vdW confinement effect. Moreover, bottom-up manufacture is an alternative feasible method to fabricate such structures [29, 30].

In conclusion, we proposed a new mechanism, van der Waals confinement, to enhance phonon transport. The vdW confinement originates from the potential barriers formed by ordered vdW interactions, which leads to a confinement of atomic thermal vibrational displacement. Thus, the vdW confinement could modulate phonon transport and the propagation of other energy carriers. An alternative artificial CPL structure was established based on vdW confinement. The EMD simulation results show that the thermal conductivities of CPL structures for PE and GNR systems at room temperature (181 $Wm^{-1}K^{-1}$ and 254 $Wm^{-1}K^{-1}$) are two-fold and three-fold as those of SC-PE and SL-GNR, respectively. Morphology analyses exhibit that the atomic thermal vibration is barrier confined, which means the reduction of anharmonicity in the system. Through mapping the potential energy, it is found that the CPL structures present confined vdW potential barriers, namely strong vdW confinement, making the atomic thermal vibrations more crystal-like. Furthermore, the CPL structures have excellent thermal properties along two directions, which goes a step further in the enhancement of thermal transport and extends the scope of applications. Polymers with high thermal conductivity may also have other technological advantages such as low cost, light weight, electrical insulation and chemical stability, so our work provides a promising alternative heat transfer materials to solve the crucial heat dissipation issue which is the major reason for the breakdown of micro/nano-electronic devices. We studied how vdW interactions can modulate the transport of heat carriers/phonons, which will be of great general interest to both experimentalists and theorists in a broad field, such as phononics [50], thermoelectrics [51, 52], and multi-carrier nonequilibrium dynamics [53, 54]. Moreover, the vdW confinement giving rise to the recovery of crystal-like atomic vibrations may weaken the electron-phonon scattering thus boost the electrical transport [55, 56], which is important for polymer-like soft materials in flexible electronics. Additionally, the photoelectric properties of soft materials in organic solar cells and light emitting diodes, might also be influenced by the morphology [57, 58] and thus could be modulated by vdW confinement.

**Acknowledgement**

The work was supported by the National Natural Science Foundation of China No. 51576076 (N. Y.), No. 51606072 (C. D.), No. 51576077 (X. H.), No. 51711540031 (N. Y.), and Hubei Provincial Natural Science Foundation of China No. 2017CFA046 (N. Y.). The authors acknowledge stimulating discussions with Quanwen Liao and Hongru Ding. The authors thank the National Supercomputing Center in Tianjin (NSCC-TJ) and China Scientific Computing Grid (ScGrid) for providing help in computations.

# How Does van der Waals Confinement Enhance Phonon Transport?

Xiaoxiang Yu[1, 2, #], Dengke Ma[1, 2, #], Chengcheng Deng[3, *], Xiao Wan[1, 2], Meng An[1, 2], Han Meng[1, 2], Xiaobo Li[3], Xiaoming Huang[3], Nuo Yang[1, 2, *]

[1] State Key Laboratory of Coal Combustion, Huazhong University of Science and Technology, Wuhan 430074, P. R. China

[2] Nano Interface Center for Energy (NICE), School of Energy and Power Engineering, Huazhong University of Science and Technology, Wuhan 430074, P. R. China

[3] School of Energy and Power Engineering, Huazhong University of Science and Technology, Wuhan 430074, P. R. China

# X. Y. and D. M. contributed equally to this work.

* To whom correspondence should be addressed. E-mail: nuo@hust.edu.cn (N.Y.), dengcc@hust.edu.cn (C.D.)

1. Simulation structures of CPL-PE and CPL-GNR.

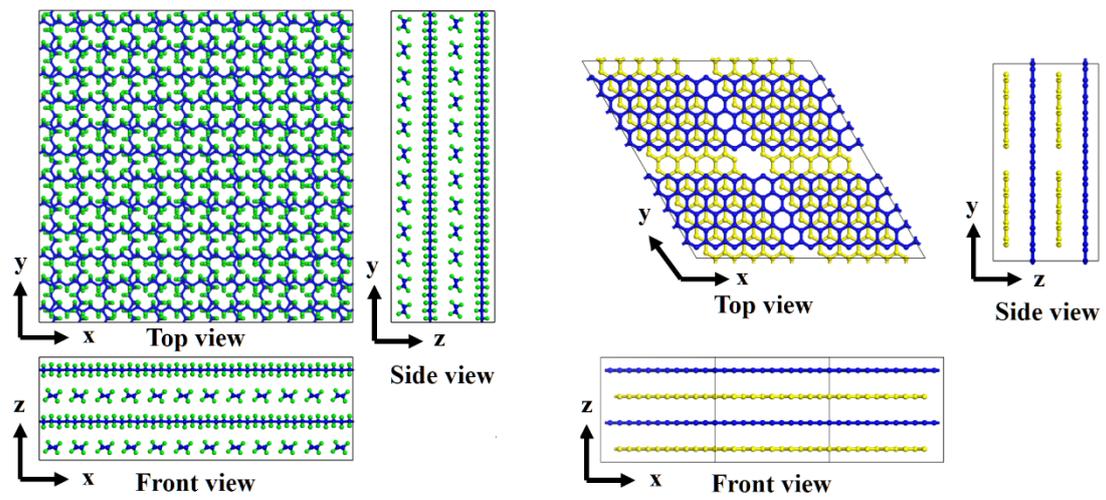

FIG. S1. Three views of CPL-PE and CPL-GNR.

2. Relaxation in NPT ensemble

To get the relaxed CPL-PE structure, initial structure is minimized by standard conjugate-gradient energy-minimization method in LAMMPS. Then, the system runs in isobaric-isothermal ensemble (NPT) for 100 ps. Fig. S2 shows the stress as a function of time, indicating no stress in all three directions after NPT process. For CPL-PE simulation system, the average system size after relaxation is 5.16 nm × 3.42 nm × 5.16 nm. Its chain is 20-unit-cell long with initial length of 5.08 nm, which denotes a slight tensile strain of 1.6% that almost has no influence on the TC of PE chain [3].

As for the thermodynamic stability, we agree that simulation system should run for long enough time (far beyond MD time scale) to achieve a stable state. In this work, we performed NPT relaxation process and obtained a reasonable mass density (0.975 g/cm$^3$). Moreover, we conducted the MD simulation at wide temperature range (250K – 400K) and the CPL-PE structure still keeps stable at high temperature (400K).

The CPL structures are probably fabricated according to a recent experimental research on the weaving of organic threads [4]. Previous experiments also reported many analogous structures such as orthogonal self-assembly multilayer polymer meshes [5] and ordered polymer structure [6], which exemplified that such kind of structures could be thermodynamically stable.

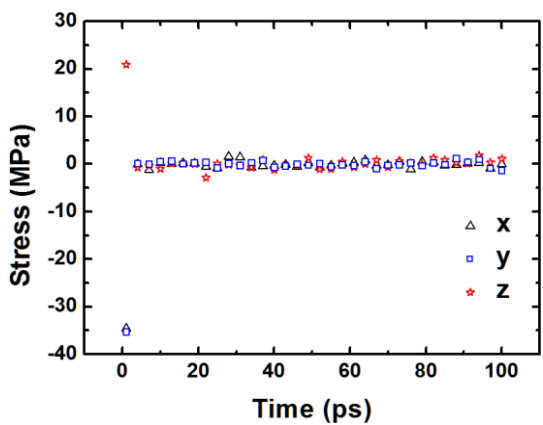

FIG. S2. Stress of CPL-PE system during NPT relaxation process.

3. MD simulation details.

Table S1. Parameter settings in MD simulation.

| Method | Equilibrium MD (Green-Kubo method) | | | | |
|---|---|---|---|---|---|
| **Potential** | | | | | |
| Type | AIREBO | | | | |
| **Simulation process** | | | | | |
| Ensemble | Setting | | | | Purpose |
| NPT | Time step (fs) | 0.1 | Runtime (ns) | 0.1 | Relax structure |
| | Temperature (K) | 300 | Pressure (atm) | 0 | |
| | Boundary condition | X, Y, Z: periodic, periodic, periodic | | | |
| NVT | Time step (fs) | 0.1 | Runtime (ns) | 0.01 | Relax structure |
| | Temperature (K) | 300 | Thermostat | Nose Hoover | |
| | Boundary condition | X, Y, Z: periodic, periodic, periodic | | | |
| NVE | Sample interval time (fs) | 1 | Runtime (ns) | 5 | Record trajectory |
| | Correlation time (ps) | 500 | Temperature (K) | 300 | |
| | Boundary condition | X, Y, Z: periodic, periodic, periodic | | | |
| **Recorded physical quantity** | | | | | |
| Temperature | $<E> = \sum_{i=1}^{N} \frac{1}{2} m v_i^2 = \frac{1}{2} N k_B T_{MD}$ | | | | |
| Heat flux | $\vec{J}(\tau) = \sum_i \vec{v}_i \varepsilon_i + \frac{1}{2} \sum_{i,j} \vec{r}_{ij} (\vec{F}_{ij} \cdot \vec{v}_i)$ | | | | |
| Thermal conductivity | $\kappa = \frac{1}{3 k_B T^2 V} = \int_0^\infty <\vec{J}(\tau) \cdot \vec{J}(0)> d\tau$ | | | | |

In all simulations, The velocity Verlet algorithm [1] is employed to integrate equations of motion. The initial conditions for all simulations use the conjugate gradient minimized equilibrium positions and random velocities corresponding to given temperatures. The system runs in isobaric–isothermal ensemble at 0 bar for 100 picosecond with different initial conditions (see Fig. S1). We used a 0.1 femtosecond time step and recorded the trajectories every 1 femtosecond at equilibrium state in microcanonical ensemble for 5 nanosecond with total energy mean variance of 0.005%. The Green-Kubo formula relates the equilibrium fluctuations of heat current, in terms of autocorrelation function, to thermal conductivity via the fluctuation-dissipation theorem. Based on the lattice parameters of ideal PE crystal, the cross-sectional-area of SC-PE is set as 18 Å$^2$ [2]. We used a combination of time and ensemble sampling to obtain a better average statistics. The results presented are averaged over 5 independent simulations with different initial conditions.

4. Chain length dependence of κ

In order to get the appropriate system length for EMD simulations, the size dependence of thermal conductivity of SC-PE, Array-PE, CPL-PE and CPL-GNR at 300 K are presented in Fig. S3. Moreover, $\kappa_{SC-PE}$, $\kappa_{Array-PE}$, $\kappa_{CPL-PE}$ and $\kappa_{CPL-GNR}$ almost converge with chain length longer than 5nm. Therefore, 5 nm length is selected to overcome the finite size effect.

There is two widely used molecular dynamics simulation methods, non-equilibrium molecular dynamics (NEMD) and equilibrium molecular dynamics (EMD). In our manuscript, EMD is preferred, instead of NEMD used in Ref. 12 [Phys. Rev. B 86, 104307 (2012)].

The advantage of EMD is to simulate a 1D structure (nanotube/single-chain) with an infinite length or a bulk structure [7-9], where a periodic boundary conditions is used. That is, the phonons will not be scatted at the boundaries. When the simulation cell is large enough, the calculated value of thermal conductivity converges due to a competing effect of increasing on both phonon modes and scatterings [10-12]. In our work, Fig. S3 shows that thermal conductivity of single chain converges and is almost independent of the simulation length when the simulation cell is larger than 5 nm. Liao et al. also reported [13] a similar result of PE single chain by EMD.

However, the EMD is not good in study a nanostructure with a finite length where the size effect is important. The phonons are scattered at boundaries, which confines phonon mean free paths. In such cases, it is well simulated by NEMD method [12, 14]. Ref. 12 [Phys. Rev. B 86, 104307 (2012)] reported that thermal conductivity of single PE chain increased with its length by NEMD method with a fixed boundary conditions.

More details about EMD and NEMD can be found in Ref. [11, 12].

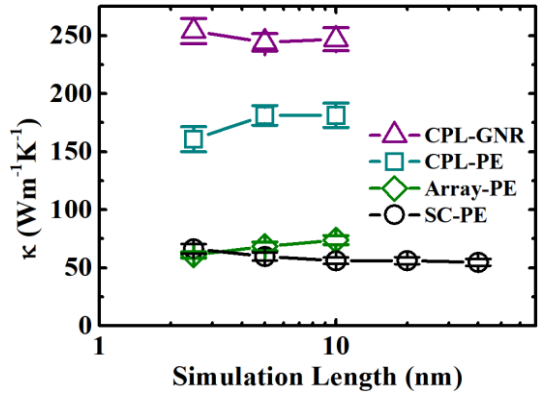

FIG. S3. Size dependence of thermal conductivity of SC-PE, Array-PE, CPL-PE and CPL-GNR.

## 5. Radial distribution function (RDF)

The reference atom ($x_0, y_0, z_0$) is the average coordinates of all atoms. We select a chain along $x$ direction in CPL-PE. RDF is calculated according to the formula of $g(r) = N/(2n-1)$. The distance ($r$) is defined as $r = [(y-y_0)^2 + (z-z_0)^2]^{1/2}$. $N$ is the count of atoms in $n$th annulus ($R \leq r < R + d$). $A(r)$ is the area of annulus. The width of each annulus ($d$) is set to 0.2 Å. The value of $\pi d^2$ is normalized as 1, thus the area of $n$th annulus is $A(r) = \pi(nd)^2 - \pi[(n-1)d]^2 = (2n-1)\pi d^2$.

6. Local vdW potential barrier of a PE chain in CPL-PE and Array-PE

    Firstly, we move a PE chain along z axis in CPL-PE (Array-PE) translationally for a distance of 0.01 Å per step within a range of 2 Å × 2 Å in xy plane and calculate the potential energy of the PE chain. Then we rotate the same PE chain while keeping other chains static to attain the angle dependence of potential energy change.

7. Potential barriers in GNR systems

We calculate the potential barriers to elaborate the vdW confinement in GNR systems. As shown in Fig. S4, the mapping potential of CPL-GNR shows slight shrink from center to edge along z direction while that of Array-GNR does not. It means that, compared with array, CPL exhibits slightly stronger confinement along z direction, thus leads to slightly larger thermal conductivity.

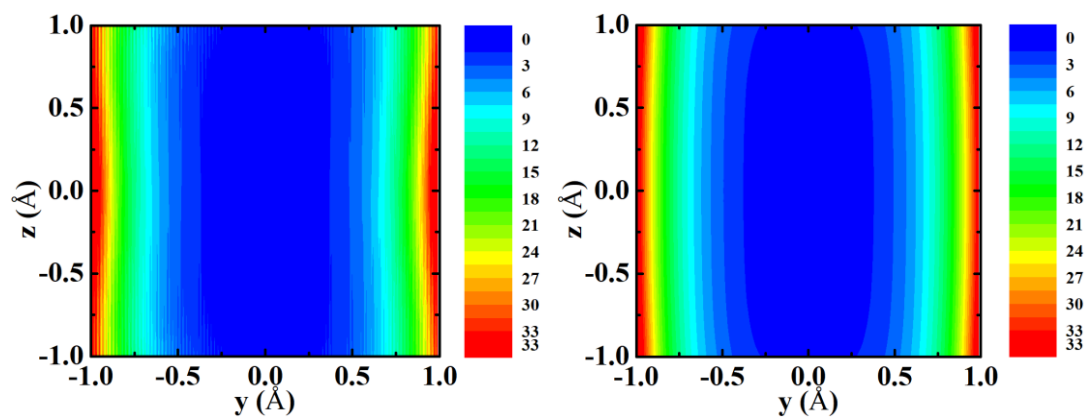

FIG. S4. The potential barrier felt by GNR in array (left) and CPL (right).